\documentclass{article}
\usepackage[english]{babel}
\usepackage{graphicx}
\usepackage{amssymb,amsmath,fullpage,subfigure}
\usepackage{amsfonts,floatrow,wrapfig,multirow,relsize}
\usepackage{amsfonts,amsmath,amsthm,array,amssymb}
\usepackage{hyperref}
\usepackage{color}
\usepackage{xcolor}
\usepackage{mathrsfs}
\usepackage[english]{babel}

\include{latexsymb}
\include{pxfonts}
\include{displaymath,enumerate}
\include{tipa}

\begin{document}

 \begin{center}
 {\Large \textbf{On the Ristic-Balakrishnan distribution:
 bivariate extension and characterizations}}
 \end{center}
 \bigskip
 \begin{center}
 {\bf \large Indranil Ghosh}\\ {\large University of North Carolina, Wilmington, North Carolina}\\
 e-mail: {\it  ghoshi@uncw.edu}\\
 \mbox{}\\
 \mbox{}\\
  {\bf \large GG Hamedani}\\ {\large Marquette University, Milwaukee, Wisconsin}\\
 e-mail: {\it gholamhoss.hamedani@marquette.edu}
 \end{center}

\begin{abstract}
Over the last few decades, a significant development has been made towards
the augmentation of some well-known lifetime distributions by various
strategies. These newly developed models have enjoyed a considerable amount
of success in modeling various real life phenomena. Motivated by this,
Ristic \& Balakrishnan (2012) developed a special class of univariate
distributions (see Ristic- Balakrishnan (2012)). Henceforth we call this
family of distribution as RB-$G$ family of distributions. The RB-$G$ family
has the same parameters of the G distribution plus an additional positive
shape parameter $a$. Several RB-$G$ distribution can be obtained from a
specified $G$ distribution. For $a = 1$, the baseline $G$ distribution is a
basic exemplar of the RB-G family with a continuous crossover towards cases
with various shapes. In this article we focus our attention on the
characterization of this family and discuss some structural properties of
the bivariate RB-$G$ family of distributions which are not discussed in
detail in Ristic and Balakrishnan (2012).
\end{abstract}

\section{ Introduction}

\bigskip

The statistics literature is filled with hundreds of continuous univariate
distributions. In the last two decades, considerable amount of work has been
done on introducing various univariate and bivariate non-normal models and
then discussing their properties, fit and applications; for elaborate
details, one may refer to the books by Kotz, Balakrishnan and Johnson (2000)
and Balakrishnan and Lai (2009). There has been an increased interest in
defining new generated classes of univariate continuous distributions by
introducing additional shape parameters to the baseline model motivated by
the need to fit various observed phenomena, specially in those situations,
where the baseline probability distribution fails to fit them adequately.
One such model that has been studied extensively in the literature is the
Ristic- Balakrishnan $G$ family of distributions ( henceforth RB in short); see Ristic- Balakrishnan
(2012) for pertinent details.

\bigskip

The RB- $G(a)$ (where $a>0$ is the parameter) pdf (probability density
function), cdf (cumulative distribution function) and the hazard function
are given, respectively, by

\begin{equation}
f(x)=\frac{1}{\Gamma(a)}\left(-\log G(x)\right)^{a-1}g(x),\quad x\in \mathbb{%
R},
\end{equation}

\noindent
\begin{equation}
F(x)=1-\frac{\int_{0}^{-\log G(x)}t^{a-1}\exp (-t)dt}{\Gamma (a)}=1-\frac{%
\gamma \left( a,-\log G(x)\right) }{\Gamma (a)},\text{ \ \ }x\in
\mathbb{R}
,
\end{equation}

\noindent and

\begin{equation}
h(x)=\frac{\left( -\log G(x)\right) ^{a-1}g(x)}{\gamma \left( a,-\log
G(x)\right) },
\end{equation}

\noindent where $G(x)$ is any baseline cdf and $\gamma (a,x)$ is the upper
incomplete gamma function.

\bigskip

An extensive survey on the univariate RB-$G$ model is given in Cordeiro et
al. (2015) while the bivariate and subsequently multivariate generalization
of such gamma-generated models are discussed in  (Ristic and Balakrishnan
(2016)). In this work we focus mainly on the characterizations of the
univariate RB-$G$ family via hazard function, moments of truncated order
statistics and many other strategies. Further, we consider conditional specification
approach to construct a class of bivariate RB-$G$ type distributions in
which both of the conditional distributions (i.e., $Y$ given $X=x$ and $X$
given $Y=y$) belong to the univariate RB-$G$ family with appropriate
parameters.  The major objective of this article is two fold: one is to characterize the univariate RB-$G$ family and the other is to provide a bivariate extension of such a family, considering 
the situation in which both the conditionals belong to the univariate RB-$G$ distribution with appropriate parameters.

\bigskip

The article is outlined as follows: In section 2 we discuss the construction
and characterization of bivariate RB-$G$ family of  distributions and discuss
some stochastic properties of the assumed model. Section 3 represents one
type of characterization via Lorenz ordering. In section 4 we discuss the
closure property of the RB-$G$ family of distributions via sample extremum.
In section 5 we consider characterizations based on two truncated moments.
Section 6 represents characterization based on truncated moment of the first
order statistic. Characterization of RB-$G$ distribution in terms of  hazard
function is presented in section 7. {\color{red}Some discussion on the estimation of the model parameters via the method of maximum likelihood are discussed in section 8.}  Finally, some concluding remarks are made in section 9.

\section{Characterization via conditional specification approach}

Let us suppose that the random variable $X$ for given $Y=y$ for each fixed $%
Y=y$ is distributed as RB-$G$ with parameter $\delta(y)$ and the parent
distribution $G$ and the random variable $Y$ for given $X=x$ for each fixed $%
X=x$ is distributed as RB with parameter $\psi(x)$ and the parent
distribution $G$. Also, let $f(x)$ and $f(y)$ be the marginal distributions
of the random variables $X$ and $Y$, respectively. Then the joint
distribution $f(x,y)$ of the random variables $X$ and $Y$ can be written as
\begin{equation*}
f(x,y)=f(x)\left[ \frac{1}{\Gamma (\psi(x))}\left( -\log G(y)\right)
^{\psi(x)-1}g(y)\right] =f(y)\left[ \frac{1}{\Gamma (\delta(y))}\left( -\log
G(x)\right) ^{\delta(y)-1}g(x)\right] .
\end{equation*}%

\noindent Our conditional density of $X$ given $Y=y$ can be rewritten in the following form

\begin{equation}
  f(x|y)=\left[\Gamma (\delta(y))\right]^{-1}\left[-\log G(x)\right]^{-1}\exp\left[\delta(y)\log\left(-\log G(x)\right)+\log g(x)\right].
  \end{equation}
\noindent This can be expressed in the form of $\ell_{1}=2$ parameter family of densities (Ref. Definition 4.1, Arnold et al. (1999)) of the form

$$f(x|y)=r_{1}(x)\beta_{1}\left(\vec{\theta(y)}\right)\exp\left(\sum_{i=1}^{2}\theta_{i}(y)q_{1i}(x)\right).$$

In our case, we have the following:

\begin{itemize}
\item $r_{1}(x)=\left(\log \left[-\log G(x)\right]\right)^{-1},$
{\color{red}\item $\beta_{1}\left(\vec{\theta(y)}\right)=\left[\left[\Gamma (\delta(y))\right]^{-1},  \quad \quad 1\right],$}
\item $\theta_{1}(y)=\delta(y),$  \quad \quad $\theta_{2}(y)=1,$
\item $q_{11}(x)=\log\left(-\log G(x)\right),$ \quad $q_{12}(x)=1.$
\end{itemize}

Similarly, the other conditional density can be rewritten in the same form but with different parametric
configuration and replacing x by y in appropriate places. If the above holds true, then according to Theorem
4.1, of Arnold et al. (1999) the bivariate density $f(x, y)$ will be of the form

\begin{equation}
  f(x,y)=r_{1}(x)r_{2}(y)\exp\left[\left(\underline{q_{1}}(x)\right)^{T} M  \quad \underline{q_{2}}(y)\right],
  \end{equation}
  
  \noindent where $T$ stands for transpose. Also, $$\underline{q_{1}}^{T}(x)=\left(q_{10}(x) \quad \quad q_{11}(x) \quad \quad q_{12}(x)\right),$$
  
  and 
  {\color{red}
  $$\underline{q_{2}}^{T}(y)=\left(q_{20}(y) \quad \quad q_{21}(y) \quad \quad q_{22}(y)\right),$$}
  
  where $q_{10}(x) = q_{20}(y) = 1$ and {\color{red}  $M$  is a  $3 \times 3$ matrix of
constants subject to} $\int_{0}^{\infty}\int_{0}^{\infty}f(x,y)dxdy=1.$ ((Ref. Equation 4.5 of Arnold et al. (1999), page 76).
In our case we can write $M$ as

\[ M=\begin{bmatrix}
m_{00} & m_{01} & m_{02}\\
m_{10} & m_{11} & m_{12}\\
m_{20} & m_{21} & m_{22}
\end{bmatrix}.
\]

Also, in our case  $$\underline{q_{1}}(t)=\underline{q_{2}}(t)= \begin{bmatrix}
           1 \\
            \log\left(-\log G(t)\right)\\
                      \log g(t)
         \end{bmatrix}.
         $$

Then from (2), we can write the joint density $f(x, y)$ as follows:
  
\begin{equation}
\begin{split}
f(x,y)&=\left[\log \left[-\log G(x)\right]\right]^{-1}\left[\log \left[-\log G(y)\right]\right]^{-1}
\exp \Bigl(m_{00}-m_{01}\log\left(\log G(y)\right)+ m_{02}\log g(y)\\
&-m_{10}\log\left(\log G(x)\right)+m_{11}\log\left(\log G(x)\right)\log\left(\log G(y)\right)-m_{12}\log\left(\log G(x)\right)\log g(y)\\
&-m_{20}\log g(x)- m_{21}\log\left(\log G(y)\right)\log g(x)+m_{22}\log g(x)\log g(y)\Bigr).
\end{split}
\end{equation}

\noindent Observe that for model (6), independence will be achieved if the following holds true: $m_{11} = m_{12} =
m_{21} = m_{22} =0$ and $m_{11} > 0, \quad m_{20} > 0, \quad m_{01} > 0, \quad m_{02} > 0.$
Note that sometimes the joint density might lead us to some nonstandard models (in the sense that they
might have a valid joint density but might not produce valid marginal densities and vice versa). Hence, we
do need appropriate constraints on the choice of $m_{ij}, i, j =1,2.$

\bigskip

\textbf{Some observations:}

\begin{itemize}
\item For various choices of $\delta(y)$ and $\psi(x)$ functions, one can
obtain various bivariate probability distribution models.

\item Note that in the above model, $m_{11} = m_{12} =
m_{21} = m_{22} =0$ implies independence.

\item From the elements of the matrix $M,$ one can establish the following relationships among the elements $m_{ij}, i, j =1,2$ which will indicate whether we will have positive {\color{red}or} negative dependence. The
simple way to look at it is by the expression of the determinant of the matrix $M,$ which is
$$|M|=m_{00}\left(m_{11}m_{22}- m_{12}m_{21}\right)- m_{01}\left(m_{10}m_{22}- m_{12}m_{20}\right)+m_{02}\left(m_{10}m_{21}- m_{11}m_{20}\right).$$

From this, we can say the following (with the assumption that all $m_{ij} > 0$):
\begin{itemize}
\item One will have positive dependence iff $\frac{m_{22}}{m_{12}}<\frac{m_{20}}{m_{10}}<\frac{m_{21}}{m_{11}}.$
\item One will have negative dependence iff $\frac{m_{22}}{m_{12}}>\frac{m_{20}}{m_{10}}>\frac{m_{21}}{m_{11}}.$
\end{itemize}

\item If $m_{12}>0$ and $m_{21}>0$ and $m_{22}\leq 0$ , then still (6) is a
 legitimate joint density. However, if any of $m_{12}$ and/or $%
m_{21}$ is 0, then the model is improper in the sense that it is no longer
integrable, although nonnegative.

\item If both $\delta(y)$ and $\psi(x)$ are linear functions of $y$ and $x$
respectively, for example, {\color{red}$\delta(y)=a_{01}+a_{02}y$} and $%
\psi(x)=a_{03}+a_{04}x$ with the condition that \ $a_{02}>0$ $,a_{04}>0$,
and $a_{01}=a_{02}\neq 0$, then we get the joint density of an exponential
family of {\color{red}distributions.} However, if those restrictions are replaced by other
possibilities, we might get the joint density for a truncated exponential.

\item If $\delta(y)=\left( \frac{y}{\sigma _{1}}\right) ^{2}+1$ and $%
\psi(x)=\left( \frac{y}{\sigma _{1}}\right) ^{2}+1$, for some non-negative
constants $\sigma _{1}$, \quad $\sigma _{2}$, then (6) will produce one of
those models which Bhattacharya (1943) identified as nonstandard models with
normal conditional distributions.

\item If $\delta(y)=b_{0}y^{2}+1$ and $\psi(x)=c_{0}x^{2}+1$, where $b_{0}>0$
and $c_{0}>0$ are some constants, then $f(x,y)$ will produce a bivariate
distribution with normal conditionals. 

Since both of the conditionals are in Gamma family and can be written (we already have used that
representation) in the form {\color{red} of }Equation (4.32) of Arnold et al. (1999), then the joint density is of the
form (Equation 4.33, page 83, Arnold et al. (1999)) we will have the following:

\begin{enumerate}

\item The conditional distribution of $X$ given $Y = y$ is

$$X|Y=y\sim \text{Gamma}\left(m_{20}+m_{22}\log g(y)+m_{21}\log\left(-\log G(y)\right), m_{10}+ m_{11}\log\left(-\log G(y)\right)+m_{12}\log g(y)\right).$$

 \noindent Similarly, the conditional distribution of $Y$ given $X= x$ is

$$Y|X=x\sim \text{Gamma}\left(m_{02}+m_{22}\log g(x)+m_{12}\log\left(-\log G(x)\right), m_{01}+ m_{11}\log\left(-\log G(x)\right)+m_{21}\log g(x)\right).$$

\bigskip

\noindent  Consequently, using known results for gamma distribution we may verify that the general $k$-th order moment ($k\geq 2$) will be

\begin{align*}
&E\left(Y^{k}|X=x\right)\notag\\
&=\frac{\left(m_{20}+m_{22}\log g(y)+m_{21}\log\left(-\log G(y)\right)+k-1\right)\cdots \left(m_{02}+m_{22}\log g(x)+m_{12}\log\left(-\log G(x)\right)\right)}
{\left(m_{10}+ m_{11}\log\left(-\log G(y)\right)+m_{12}\log g(y)\right)^{k}}.
\end{align*}

An analogous expression for the $k$-th order conditional moment for the other conditional distribution can be easily obtained.

\item The marginal density of $X$ will be

\begin{eqnarray*}
f_{X}(x)
&=&\left[\log \left[-\log G(x)\right]\right]^{-1}\frac{\Gamma\left(m_{02}+m_{22}\log g(x)+m_{12}\log\left(-\log G(x)\right)\right)}{\left[m_{01}+ m_{11}\log\left(-\log G(x)\right)+m_{21}\log g(x)\right]^{m_{02}+m_{22}\log g(x)+m_{12}\log\left(-\log G(x)\right)}}\notag\\
&&\times \exp\left(m_{00}+m_{10}\log\left(-\log G(x)\right)+m_{20}\log g(x)\right), x>0.
\end{eqnarray*}

Similarly, one can find an analogous expression for the density of $Y.$
\end{enumerate}
 
\bigskip

\subsection{Some distributional properties}
\item  \textbf{Shape of the distribution:} A critical point of a function with two variables is a point where the partial
derivatives of first order are equal to zero. There are two reasons as to why it is important to find
the critical points of a bivariate probability distribution: (a) To determine the shape of the distribution in
order to find {\color{red}it's} flexibility in fitting a data which is exhibiting a similar shape pattern, and (b) To
identify the number of and the locations of modes of the density. In order to identify the location of the
mode of the density (6), we consider the first derivatives of  $\log f(x, y)$ with respect to $x$ and $y$ and then equate to zero.  This results in  the following two equations:

\begin{align*}
&\frac{\partial f(x,y)}{\partial x}
=\left(-\log G(x)\right)^{-1}\frac{g(x)}{G(x)}\left[-\left(\log\left(-\log G(x)\right)\right)^{-2}-m_{10}+ m_{11}\log\left(\log G(y)\right)-m_{12}\log g(y)\right]\\
&+\frac{g('x)}{g(x)}\left[-m_{20}- m_{21}\log\left(\log G(y)\right)+m_{22}\log g(y)\right]
=0.
\end{align*}

Also,
\begin{align*}
&\frac{\partial f(x,y)}{\partial y}
=\left(-\log G(y)\right)^{-1}\frac{g(y)}{G(y)}\left[-\left(\log\left(-\log G(y)\right)\right)^{-2}-m_{01}+ m_{11}\log\left(\log G(x)\right)-m_{21}\log g(x)\right]\\
&+\frac{g('y)}{g(y)}\left[-m_{02}- m_{12}\log\left(\log G(x)\right)+m_{22}\log g(x)\right]
=0.
\end{align*}

\noindent It is clear from the above that for a baseline $G$ distribution a numerical evaluation is required as analytical expressions are difficult to obtain.

\item  Next, we focus our attention to some dependence
properties of the bivariate distribution in (6). There are various ways to
describe and measure the dependence or association between two random
variables. A distribution is said to be positive likelihood ratio dependent
(PLRD) if its pdf $f(x,y)$ satisfies $\frac{f(x_{1},y_{1})f(x_{2},y_{2})}{%
f(x_{1},y_{2})f(x_{2},y_{1})}\geq 1$, $\forall x_{1}>x_{2}\quad \text{and}%
\quad y_{1}>y_{2}$. The quantity $\frac{f(x_{1},y_{1})f(x_{2},y_{2})}{%
f(x_{1},y_{2})f(x_{2},y_{1})}\geq 1$ measures "local" positive (or negative)
likelihood ratio dependence at each point $(x,y)\in
\mathbb{R}
^{2}$, and its integral over the portion of $%
\mathbb{R}
^{4}$, where $x_{1}>x_{2}\quad \text{and}\quad y_{1}>y_{2}$ is a measure of
"average" likelihood ratio dependence. For the bivariate density in (6), the
above condition reduces to

\begin{equation}
c_{22}\left[ \left( a_{2}(x_{1})-a_{2}(x_{2})\right) \left(
a_{1}(y_{1})-a_{1}(y_{2})\right) \right] \times \left[ \frac{\log G(x_{1})}{%
\log G(x_{2})}\right] ^{a_{1}(y_{1})-a_{1}(y_{2})}\times \left[ \frac{\log
G(y_{1})}{\log G(y_{2})}\right] ^{a_{2}(x_{1})-a_{2}(x_{2})}\geq 1.
\end{equation}

\item \noindent Next, since $x_{1}>x_{2}$, $y_{1}>y_{2}$ if both $a_{1}(.)$
and $a_{2}(.)$ are monotonically increasing functions then (7) holds
(provided $c_{22}\geq 0$). Hence the bivariate density in (6) will exhibit
PLRD property.

This PLRD property of the density (6) implies the following:

\begin{itemize}
\item $P(X\leq x|y=y)$ is non-increasing in $y$ for all $x$,

\item $P(Y\leq y|X=x)$ is non-increasing in $x$ for all $y$,

\item {\color{red} $P(Y>y|X>x)$ is non-decreasing in $x$ for all $y$,}

\item $P(Y\leq y|X\leq x)\geq P(Y\leq y)P(X\leq x)$,

\item $P(Y>y|X>x)\geq P(Y>y)P(X>x)$.
\end{itemize}
\end{itemize}

\section{Characterization via generalized Lorenz ordering}

\bigskip

The expression for generalized Lorenz ordering ( henceforth in short GL) is given by (for a random
variable $X$) by $GL_{X}(p)=\int_{0}^{p}F_{X}^{-1}(t)dt,$ for $p\in (0,1)$.
Also, we mention here a result by Ramos et al. (2000) which is as follows:

\bigskip

If $Z_{i}\sim gamma(\alpha_{i}, \beta_{i})$ for $i=1,2$. Then, if $%
\alpha_{1}\leq \alpha_{2}$ and $\alpha_{1}\beta_{1}\leq \alpha_{2}\beta_{2}$%
, then $Z_{2}\leq_{GL} Z_{1}$. In our case, according to Ristic-
Balakrishnan model motivation, we consider the following:

\bigskip

Suppose $Z_{1}\sim gamma(\delta_{1},1)$ then $X=F^{-1}\left(1-\exp(-Z_{1})\right)\sim
RB(\delta_{1})$. Similarly, $Z_{2}\sim gamma(\delta_{2},1)$ then $%
X=F^{-1}\left(1-\exp(-Z_{2})\right)\sim RB(\delta_{2})$.

\bigskip

Now, if we assume $\delta _{1}\leq \delta _{2}$, then according to Ramos et
al. (2000) result and noting the fact that Lorenz ordering is preserved
under one-to-one transformation, assuming $X$ and $Y$ are one-to-one
transformation of $Z_{1}$ and $Z_{2}$, the result will hold in this case
also. In other words if $X\sim RB(\delta _{1})$ and $Y\sim RB(\delta _{2})$
, with $\delta _{1}\leq \delta _{2}$, then $Y\leq _{GL}X$. This is one type
of characterization for the RB-G family.

\section{Characterization via closure property of sample extremum}

\textit{Theorem 4.1:} The Ristic-Balakrishnan G family of distributions is
closed under minimization and maximization. In other words, for a random
sample ($i.i.d$) of size $n$ drawn from (1), we can write the following:%
\newline
\begin{equation*}
X_{1:n}\sim \frac{\Gamma (n\delta +s_{k})}{\left[ \Gamma (\delta )\right]
^{n}}RB(n\delta +s_{k}),
\end{equation*}%
and
\begin{equation*}
X_{n:n}\sim \sum_{j=0}^{n}(-1)^{j}\binom{n}{j}\frac{\Gamma (j\delta +s_{\ell
})}{\left[ \Gamma (\delta )\right] ^{j}}RB(j\delta +s_{\ell }),
\end{equation*}

\noindent where $X_{1:n}=\min_{1\leq i\leq n} X_{i}$ and $%
X_{n:n}=\max_{1\leq i\leq n} X_{i}$ are the smallest and largest order
statistics respectively.

\bigskip

\textit{Proof.} Let us consider

\begin{eqnarray}
P(X_{1:n}>x) &=& \left[P(X_{1}>x)\right]^{n}  \notag \\
&=&\left[\frac{\gamma\left(\delta, -\log G(x)\right)}{\Gamma(\delta)}\right]%
^{n}  \notag \\
&=&\left[\frac{1}{\Gamma(\delta)}\right]^{n} \left[\sum_{k=0}^{%
\infty}(-1)^{k}\frac{\left[-\log G(x)\right]^{\delta+k}}{k! (\delta+k)}%
\right]^{n}  \notag \\
&&\text{on using power series expansion of the incomplete gamma function}
\notag \\
&=&\left[\frac{1}{\Gamma(\delta)}\right]^{n} \left[\sum_{k_{1}}^{\infty}%
\cdots\sum_{k_{n}=0}^{\infty}(-1)^{s_{k}} \frac{(-\log G(x))^{s_{k}+n\delta}%
}{p_{k}}\right],  \notag \\
\end{eqnarray}

\noindent where $s_{k}=\sum_{i=1}^{n}k_{i}$ and $p_{k}=\prod_{i=1}^{n}
k_{i}! $. From (7), it is easy to show that $X_{1:n}\sim (\frac{%
\Gamma(n\delta+s_{k})}{\left[\Gamma(\delta)\right]^{n}}RB(n\delta+s_{k})$.
Similarly the other part of the theorem can be established.

\section{Characterization based on two truncated moments}

In this section we present characterizations of the RB-G distribution in
terms of a simple relationship between two truncated moments.  The results
derived here will employ an interesting theorem due to Gl\"{a}nzel (1987),
which is given below. \ The advantage of the characterizations given here is
that the cdf need {\color{red}not}  have a closed form and it is given as an integral
whose integrand is in terms of the solution of a differential equation. This
provides a bridge between probability and differential equation.

\bigskip

\textbf{Theorem 5.1. \ \ } Let \ $\left( \Omega ,\mathscr{F} ,\mathbf{P}%
\right) $ \ be a given probability space and let \ $H=\left[ a,b\right] $ \
be an interval for some \ $a<b$ \ $\left( a=-\infty \text{ },\text{ }%
b=\infty \text{ \ might as well be allowed}\right) .$ \ Let \ $X:\Omega
\rightarrow H$ \ be a continuous random variable with the distribution
function \ $F$ \ and let \ $q_{1}$ \ and \ $q_{2}$ \ be two real functions
defined on \ $H$ \ such that

\begin{center}
$\bigskip$

$\mathbf{E}\left[ q_{1}\left( X\right) \text{ }|\text{ }X\geq x\right] =%
\mathbf{E}\left[ q_{2}\left( X\right) \text{ }|\text{ }X\geq x\right] $ $%
\eta \left( x\right) ,$ \ \ \ \ \ $x\in H$ $,$
\end{center}

\noindent

\noindent is defined with some real function \ $\eta $ $.$ \ Assume that \ $%
q_{1}$ $,$ $q_{2}\in C^{1}\left( H\right) $ $,$ $\eta \in C^{2}\left(
H\right) $ \ and \ $F$ \ is twice continuously differentiable and strictly
monotone function on the set \ $H$ $.$ \ Finally, assume that the equation \
$q_{2}\eta =q_{1}$ \ has no real solution in the interior of \ $H$ $.$ \
Then \ $F$ \ is uniquely determined by the functions $q_{1}$ $,$ $q_{2}$ \
and \ $\eta $ $,$ particularly

\bigskip

\begin{equation*}
F\left( x\right) =\int_{a}^{x}C\left\vert \frac{\eta ^{\prime }\left(
u\right) }{\eta \left( u\right) q_{2}\left( u\right) -q_{1}\left( u\right) }%
\right\vert \exp \left( -s\left( u\right) \right) \text{ }du\text{ },
\end{equation*}

\noindent

\noindent where the function \ $s$ \ is \ a solution of the differential
equation \ $s^{\prime }=\frac{\eta ^{\prime }\text{ }q_{2}}{\eta \text{ }%
q_{2}\text{ }-\text{ }q_{1}}$ \ and \ $C$ \ is a constant, chosen to make \ $%
\int_{H}dF=1$ $.$

\bigskip

\textbf{Remarks 5.1. \ \ }$\left( a\right) $ \ In Theorem 5.1, the interval \ $H$
\ need not be closed. \ $\left( b\right) $ \ The goal is to have the
function \ $\eta $ \ as simple as possible. \ $\left( c\right) $ \ It is
possible to state  {\color{red}Theorem 5.1} based on two functions \ $q_{1}$ \ and \ $\eta $
\ by setting \ $q_{2}\left( x\right) \equiv 1$, however, the extra function
gives more flexibility as far as applications are concerned.

\bigskip

\textbf{Proposition 5.1. \ \ }Let \ $X:\Omega \rightarrow
\mathbb{R}
$ \ be a continuous random variable and let \ $q_{2}\left( x\right) =\left(
G\left( x\right) \right) ^{-1}$\ and \ $q_{1}\left( x\right) =q_{2}\left(
x\right) \left( -\log \left( G\left( x\right) \right) \right) ,$ for \ $x\in
\mathbb{R}
.$ {\color{red} Then the pdf as given by (1)} is is true  if and only if the
function \ $\eta $ \ defined in {\color{red}Theorem 5.1} has the form

\begin{center}
\bigskip

\begin{equation*}
\eta \left( x\right) =\frac{a}{a+1}\left( -\log \left( G\left( x\right)
\right) \right) ,\text{ \ \ }x\in
\mathbb{R}
.
\end{equation*}
\end{center}

\bigskip

Proof. \ Let \ $X$ \ have pdf \ $\left( 1\right) $ $,$ then

\bigskip

\begin{equation*}
\left( 1-F\left( x\right) \right) \text{ }\mathbf{E}\left[ q_{2}\left(
X\right) \text{ }|\text{ }X\geq x\right] =\frac{1}{a\Gamma \left( a\right) }%
\left( -\log \left( G\left( x\right) \right) \right) ^{a},
\end{equation*}

\noindent and

\begin{equation*}
\left( 1-F\left( x\right) \right) \text{ }\mathbf{E}\left[ q_{1}\left(
X\right) \text{ }|\text{ }X\geq x\right] =\frac{1}{\left( a+1\right) \Gamma
\left( a\right) }\left( -\log \left( G\left( x\right) \right) \right) ^{a+1},
\end{equation*}

\noindent and finally

\begin{equation*}
\eta \left( x\right) q_{2}\left( x\right) -q_{1}\left( x\right) =-{\color{red}\frac{1}{a+1}}%
q_{2}\left( x\right) \left( -\log \left( G\left( x\right) \right) \right) <0,%
\text{ \ \ }x\in
\mathbb{R}
\text{ }.
\end{equation*}

\bigskip

Conversely, \ if \ $\eta$ \ is given as above, then

\bigskip

\begin{equation*}
s^{\prime }\left( x\right) =\frac{\eta ^{\prime }\left( x\right) \text{ }%
q_{2}\left( x\right) }{\eta \left( x\right) \text{ }q_{2}\left( x\right)
-q_{1}\left( x\right) }=\frac{ag\left( x\right) }{G\left( x\right) \left(
-\log \left( G\left( x\right) \right) \right) },\text{ \ \ }x\in
\mathbb{R}
\text{ },
\end{equation*}

\noindent and hence

\begin{equation*}
s\left( x\right) =-a\log (\left( -\log \left( G\left( x\right) \right)
\right) ),\text{ \ \ \ }x\in
\mathbb{R}
.
\end{equation*}

\bigskip

\noindent Now, in view of Theorem 5.1, \ $X$ \ has cdf\ $\left( 2\right) $ \
and pdf \ $\left( 1\right) $ $.$

\bigskip

\textbf{Corollary 5.1. \ }Let \ $X:\Omega \rightarrow
\mathbb{R}
$ \ be a continuous random variable and let \ $q_{2}\left( x\right) $ \ be
as in Proposition 5.1 $.$ \ {\color{red}Then the pdf as given by (1)} is is valid if and only  if
 there exist functions \ $q_{1}$ \ and \ $\eta $ \ defined in
Theorem $5.1$ satisfying the differential equation

\bigskip

\begin{equation*}
\frac{\eta ^{\prime }\left( x\right) q_{2}\left( x\right) }{\eta \left(
x\right) q_{2}\left( x\right) -q_{1}\left( x\right) }=\frac{ag\left(
x\right) }{G\left( x\right) \left( -\log \left( G\left( x\right) \right)
\right) },\text{ \ \ }x\in
\mathbb{R}
\text{ }.
\end{equation*}

\bigskip

\textbf{Remarks 5.2. \ }$\left( a\right) $ The general solution of the
differential equation in Corollary1 is

\bigskip

\begin{equation*}
\eta \left( x\right) =\left( -\log \left( G\left( x\right) \right) \right)
^{a}\left[ -\int a\frac{g\left( x\right) }{G\left( x\right) }\left( -\log
\left( G\left( x\right) \right) \right) ^{a-1}\left( q_{2}\left( x\right)
\right) ^{-1}q_{1}\left( x\right) dx+D\right] ,\text{ \ }x\in
\mathbb{R}
,
\end{equation*}

\bigskip

\noindent where \ $D$ \ is a constant. \ One set of appropriate functions is
given in Proposition 5.1 with \ $D=0.$

$\left( b\right) $ \ Clearly there are other triplets of functions \ $\left(
q_{1},q_{2},\eta \right) $ satisfying the conditions of Theorem 5.1. \ We
presented one such triplet in Proposition1.

\bigskip

\section{Characterizations based on truncated moment of the $1st$ \textbf{%
order statistic}}

\bigskip

Let \ $X_{1:n}\leq X_{2:n}\leq ...\leq X_{n:n}$ \ be the corresponding order
statistics from a random sample of size $n$ of a continuous cdf $F.$ We
briefly discuss here characterization results based on functions of the \ $%
1st$ \ order statistic. \ We like to mention here that the proof of
Proposition 2 below is straightforward extension of that of \ Theorem 2.2 of
Hamedani (2010). \ We give a short proof of it for the sake of completeness.

\bigskip

\textbf{Proposition 6.1.} \ \textbf{\ }Let \ $X:\Omega \rightarrow
\mathbb{R}
$ \ be a continuous random variable with cdf \ $F$ . \ Let \ $\psi \left(
x\right) $ \ and \ $q\left( x\right) $ \ be two differentiable functions on $%
\mathbb{R}
$ \ such that

\bigskip

\begin{equation*}
\lim_{\text{ }x\rightarrow \infty }\psi \left( x\right) \left[ 1-F\left(
x\right) \right] ^{n}=0\ ,\text{ \ \ \ }\ \int_{-\infty }^{\infty }\frac{q%
\text{ }^{\prime }\left( t\right) }{\left[ q\left( t\right) -\psi \left(
t\right) \right] }dt=\infty .
\end{equation*}

\bigskip\ Then

\bigskip\ \

\begin{equation}
E\left[ \psi \left( X_{1:n}\right) |\text{ }X_{1:n}>t\right] =q\left(
t\right) \text{ },\text{ \ }t\in
\mathbb{R}%
\end{equation}

\bigskip

\noindent implies

\begin{equation}
F\left( x\right) =1-\exp \left\{ -\int_{-\infty }^{x}\frac{q^{\prime }\left(
t\right) }{n\left[ q\left( t\right) -\psi \left( t\right) \right] }%
dt\right\} ,\text{ \ \ }x\in
\mathbb{R}
.\text{ \ \ \ \ \ \ }
\end{equation}

\bigskip

\textit{Proof.} \ If \ $\left( 9\right) $ \ holds, then using integration by parts on
the left hand side of \ $\left( 9\right) $ \ and the assumption\ \ $\lim_{%
\text{ }x\rightarrow \infty }\psi \left( x\right) \left[ 1-F\left( x\right) %
\right] ^{n}=0$ , we have

\bigskip

\begin{align*}
& \int_{t}^{\infty}\psi^{\prime}\left( x\right) \left( 1-F\left( x\right)
\right) ^{n}dx \\
& =\left[ q\left( t\right) -\psi\left( t\right) \right] \left( 1-F\left(
t\right) \right) ^{n}\text{ }.
\end{align*}

\bigskip

Differentiating both sides of the above equation with respect to \ $t$ , we
arrive at

\bigskip

\begin{equation}
\frac{f\left( t\right) }{1-F\left( t\right) }=\frac{q^{\prime }\left(
t\right) }{n\left[ q\left( t\right) -\psi \left( t\right) \right] },\text{ \
\ }t\in
\mathbb{R}
\text{. \ \ \ \ \ \ \ \ \ \ \ \ \ \ \ \ \ \ \ \ \ \ }
\end{equation}

\bigskip\ \ \ \ \ \ \ \ \ \ \ \ \ \ \

Now, integrating \ $\left( 11\right) $ \ from \ $-\infty $ \ to \ $x$ , we
have, in view of \ $\int_{-\infty }^{\infty }\frac{q\text{ }^{\prime }\left(
t\right) }{\left[ q\left( t\right) -\psi \left( t\right) \right] }dt=\infty $
, a cdf $F$ \ given by \ $\left( 10\right) .$

\bigskip

\textbf{Remarks 5.3. } $\left( a\right) $ Taking, for instance, \textbf{\ }$%
\psi \left( x\right) =\left( \gamma \left( a,-\log \left( G\left( x\right)
\right) \right) \right) ^{n}$ \ and $\ q\left( x\right) =\frac{1}{2}\psi
\left( x\right) $ in Proposition 6.1, we arrive at $\left( 2\right) $. \ $%
\left( b\right) $ \ the above Proposition holds with the random variable \ $%
X $ \ in place of \ $X_{1:n}$ \ with of course appropriate conditions .

\bigskip

\section{Characterization based on hazard function}

\bigskip

It is obvious that the hazard function, $h_{F}$,\ of a twice differentiable
distribution function, $F$, satisfies the first order differential equation

\bigskip

\begin{equation*}
\frac{h_{F\text{ }}^{\prime}\left( x\right) }{h_{F\text{ }}\left( x\right) }%
-h_{F\text{ }}\left( x\right) =q\left( x\right) \text{ },
\end{equation*}

\bigskip

\noindent where \ $q\left( x\right) $ \ is an appropriate integrable
function. \ Although this differential equation has an obvious form since

\bigskip

\begin{equation}
\frac{h_{F\text{ }}^{\prime }\left( x\right) }{h_{F\text{ }}\left( x\right) }%
-h_{F\text{ }}\left( x\right) =\frac{f\text{ }^{\prime }\left( x\right) }{f%
\text{ }\left( x\right) }\text{ },\text{ \ \ \ \ \ \ \ \ \ \ }
\end{equation}

\bigskip

\noindent for many univariate continuous distributions \ $\left( 12\right) $
\ seems to be the only differential equation in terms of the hazard
function. \ The goal of the characterization based on hazard function is to
establish a differential equation in terms of hazard function, which has as
simple form as possible and is not of the trivial form \ $\left( 7\right).$ \ Here, we present a characterization of the of RB-$G$ model based on a
nontrivial differential equation in terms of the hazard function.

\bigskip

\textbf{Proposition 7.1. \ }Let \ $X:\Omega \rightarrow
\mathbb{R}
$ \ be a continuous random variable.  {\color{red}Then the pdf as given by (1)} is true if and only if its hazard function \ $h_{F}$ \ satisfies the
differential equation

\bigskip

\begin{equation}
h_{F}^{\prime }\left( x\right) -\frac{g^{\prime }\left( x\right) }{g\left(
x\right) }h_{F}\left( x\right) =g\left( x\right) \frac{d}{dx}\left\{ \frac{%
\left( -\log \left( G\left( x\right) \right) \right) ^{a-1}}{\gamma \left(
a,-\log \left( G\left( x\right) \right) \right) }\right\} ,\text{\ }x\in
\mathbb{R}
.\text{ \ }
\end{equation}

\noindent

Proof: \ {\color{red}If the pdf as given by (1)} is true, then clearly \ $\left(
13\right) $ \ holds. \ Now, if \ $\left( 13\right) $ holds, then after
dividing both sides of $\left( 13\right) $ by $\ g\left( x\right) $ , we
arrive at

\bigskip

\begin{equation*}
\frac{d}{dx}\left\{ \left( g\left( x\right) \right) ^{-1}h_{F\text{ }}\left(
x\right) \right\} =\frac{d}{dx}\left\{ \frac{\left( -\log \left( G\left(
x\right) \right) \right) ^{a-1}}{\gamma \left( a,-\log \left( G\left(
x\right) \right) \right) }\right\} ,
\end{equation*}

\noindent from which we have

\noindent

\begin{equation}
h_{F\text{ }}\left( x\right) =\frac{f\left( x\right) }{1-F\left( x\right) }=%
\frac{g\left( x\right) \left( -\log \left( G\left( x\right) \right) \right)
^{a-1}}{\gamma \left( a,-\log \left( G\left( x\right) \right) \right) }.%
\text{\ \ \ }
\end{equation}

\noindent

Integrating both sides of \ $\left( 9\right) $ \ from \ $-\infty $ \ to \ $x$
, we have

\begin{equation*}
-\log (\left( 1-F\left( x\right) \right) =-\log \left\{ \frac{\gamma \left(
a,-\log \left( G\left( x\right) \right) \right) }{\Gamma \left( a\right) }%
\right\} .
\end{equation*}

\noindent from which we obtain

\begin{equation*}
1-F\left( x\right) =\frac{\gamma \left( a,-\log \left( G\left( x\right)
\right) \right) }{\Gamma \left( a\right) },\text{ \ \ \ \ }x\in
\mathbb{R}
.
\end{equation*}

\bigskip

\textbf{Remarks 4.} \ For \ $a=2$ , equation \ $\left( 14\right) $ \ reduces
to the following simple equation

\bigskip

\begin{equation*}
h_{F}^{\prime }\left( x\right) -\frac{g^{\prime }\left( x\right) }{g\left(
x\right) }h_{F}\left( x\right) =\frac{\left( g\left( x\right) \right) ^{2}}{%
\left( \gamma \left( 2,-\log \left( G\left( x\right) \right) \right) \right)
^{2}}\left\{ \left( \log \left( G\left( x\right) \right) \right) ^{2}-\frac{%
\gamma \left( 2,-\log \left( G\left( x\right) \right) \right) }{G\left(
x\right) }\right\} \text{ },
\end{equation*}

\noindent or

\begin{equation*}
\frac{d}{dx}\left\{ \left( g\left( x\right) \right) ^{-1}h_{F\text{ }}\left(
x\right) \right\} =\frac{d}{dx}\left\{ \frac{\left( -\log \left( G\left(
x\right) \right) \right) }{\gamma \left( 2,-\log \left( G\left( x\right)
\right) \right) }\right\}
\end{equation*}

{\color{red}

\section{Estimation of model parameters}
Here we consider method of maximum likelihood of estimation under the classical approach, which we describe below:

In this case, we consider (for more general set up) the  bivariate RB-G distribution with the joint density as in (5). As pointed earlier, $m_{01}, m_{02}, m_{10}, m_{11} , m_{12}, m_{21}, m_{22}$ are constrained to make the density integrable while $m_{00}$ is evaluated, as a function of the other parameters, to make the integral equal to 1. For notational simplicity, we relabel of the model parameters by setting, $m_{01}=\theta_1, m_{02}=\theta_2, m_{10}=\theta_3, m_{11}= \theta_4, m_{12}=\theta_5, m_{20}=\theta_6, m_{21}=\theta_7, m_{22}=\theta_8$  and let

\begin{eqnarray}
\Psi(\underline{\theta})
&=&\exp\left(-m_{00}\right)\notag\\
&=&\int_{0}^{\infty}\int_{0}^{\infty}r_{1}(x)r_{2}(y)\exp\left(\theta_{1}q_{21}(x)+\theta_{2}q_{22}(y)+\theta_{3}q_{11}(x)+\theta_{4}q_{11}(x)q_{21}(y)\right.\notag\\
&&\left.+\theta_{5}q_{11}(x)q_{22}(y)+\theta_{6}q_{12}(x)+\theta_{7}q_{12}(x)q_{21}(y)+\theta_{8}q_{12}(x)q_{22}(y)\right) dx dy.
\end{eqnarray}

\smallskip

\noindent With this notation the log-likelihood of a sample of size $n$ ($(X_1,Y_1), \cdots, (X_n,Y_n)$) from our density in (5) is

\begin{align}
&\log L=\ell \notag\\
&=-n\log \Psi(\underline{\theta})+\sum_{i=1}^{n}r_{1}(X_i)+\sum_{i=1}^{n}r_{2}(Y_i)+\theta_{1}\sum_{i=1}^{n}q_{21}(X_i)+\theta_{2}\sum_{i=1}^{n}q_{22}(Y_i)+\theta_{3}\sum_{i=1}^{n}q_{11}(X_i)\notag\\
&+\theta_{4}\sum_{i=1}^{n}q_{11}(X_i)q_{21}(Y_i)
+\theta_{5}\sum_{i=1}^{n}q_{11}(X_i)q_{22}(Y_i)+\theta_{6}\sum_{i=1}^{n}q_{12}(X_i)+\theta_{7}\sum_{i=1}^{n}q_{12}(X_i)q_{21}(Y_i)+\theta_{8}\sum_{i=1}^{n}q_{12}(X_i)q_{22}(Y_i)
\end{align}
 
Differentiating and subsequently the partial derivatives equal to zero yields the following likelihood equations

\begin{equation}
\frac{\frac{\partial \Psi(\underline{\theta})}{\partial \theta_1} }{\Psi(\underline{\theta})}
=\frac{1}{n}\sum_{i=1}^{n}q_{21}(X_i)
\end{equation}

\begin{equation}
\frac{\frac{\partial \Psi(\underline{\theta})}{\partial \theta_2} }{\Psi(\underline{\theta})}
=\frac{1}{n}\sum_{i=1}^{n}q_{22}(Y_i)
\end{equation}

\begin{equation}
\frac{\frac{\partial \Psi(\underline{\theta})}{\partial \theta_3} }{\Psi(\underline{\theta})}
=\frac{1}{n}\sum_{i=1}^{n}q_{11}(X_i)
\end{equation}

\begin{equation}
\frac{\frac{\partial \Psi(\underline{\theta})}{\partial \theta_4} }{\Psi(\underline{\theta})}
=\frac{1}{n}\sum_{i=1}^{n}q_{11}(X_i)q_{21}(Y_i)
\end{equation}

\begin{equation}
\frac{\frac{\partial \Psi(\underline{\theta})}{\partial \theta_5} }{\Psi(\underline{\theta})}
=\frac{1}{n}\sum_{i=1}^{n}q_{11}(X_i)q_{22}(Y_i)
\end{equation}

\begin{equation}
\frac{\frac{\partial \Psi(\underline{\theta})}{\partial \theta_6} }{\Psi(\underline{\theta})}
=\frac{1}{n}\sum_{i=1}^{n}q_{12}(X_i)
\end{equation}

\begin{equation}
\frac{\frac{\partial \Psi(\underline{\theta})}{\partial \theta_7} }{\Psi(\underline{\theta})}
=\frac{1}{n}\sum_{i=1}^{n}q_{12}(X_i)q_{21}(Y_i)
\end{equation}

\begin{equation}
\frac{\frac{\partial \Psi(\underline{\theta})}{\partial \theta_8} }{\Psi(\underline{\theta})}
=\frac{1}{n}\sum_{i=1}^{n}q_{12}(X_i)q_{22}(Y_i)
\end{equation}

\noindent If $\Psi(\underline{\theta})$  is a simple analytic expression, the above set of equations can be easily solved (direct or iteratively). In case $\Psi(\underline{\theta})$ is very nasty in nature (which by the way is true in most cases) still can be evaluated by numerical integration. A reasonable approach (for details, see Arnold et al. (1999))  would be in terms of choosing initial values of $\theta_{j}, j=1,2,\cdots, 7$ (may be based on moment estimates) and then search for a  value of $\theta_{8}$ to make (24) valid. Next, taking this value of $\theta_{8}$ with the previous values of $\theta_{k}, k=2,3,\cdots, 7$ search for a value of $\theta_{1}$ to make (17)  valid and so on. This is without a doubt is heavily computer intensive but most likely more efficient than a direct search which might involve more numerical evaluations of $\Psi(\underline{\theta})$ for various choices of   $\underline{\theta}$. After solving the likelihood equations, we may, with the help of numerical integration, write an approximation for the variance-covariance matrix of our estimate $\hat{\underline{\theta}}$. The Fisher information matrix corresponding to our model is the $8\times 8$ matrix $I(\underline{\theta})$ with the $(i,j)$-th element given by

$$I_{i,j}(\underline{\theta})
=\frac{\Psi(\underline{\theta})\left[\frac{\partial^{2} \Psi(\underline{\theta})}{\partial \theta_{i}\partial \theta_{j}}\right]-
\left(\frac{\partial \Psi(\underline{\theta})}{\partial \theta_{i}}\right)\left(\frac{\partial \Psi(\underline{\theta})}{\partial \theta_{j}}\right)}{\left(\Psi(\underline{\theta})\right)^{2}}$$.

Then the estimated variance covariance matrix of $\hat{{\underline{\theta}}}$ is 
$\hat{\sum}\left(\underline{\theta}\right)=\frac{\left[I(\hat{\underline{\theta}}\right]^{-1}}{n},$
where $\hat{{\underline{\theta}}}$ is the solution to (17)-(24). Of course, the entries in $I(\hat{\underline{\theta}})$ must be computed numerically.  It is to be noted that for specific choices of the baseline cdf $G(.)$ there will be additional parameter choices, and that can also be evaluated with the above procedure.

}

\section{Concluding remarks}
In this paper, we discuss in brief, a bivariate extension of the univariate RB-G model and some associated
structural properties via conditional specification. Most of the structural properties for the univariate RB-G
model have been discussed in Bourguignon et al. (2016). Here we mainly focus on various useful characterizations
of the univariate RB-G model along with the bivariate extension. {\color{red} We have also introduced some inferential strategies for estimating the model parameters under the maximum likelihood method.} Inferential procedures for such
bivariate models under  the  Bayesian paradigm (for specific members of G(.)) will be the
subject matter of a different article as a future project.

\begin{center}
\textbf{References}

\bigskip
\end{center}
Arnold, B.C., Castillo, E., and Sarabia, J.M. (1999). Conditional Specification of Statistical Models.
Springer-Verlag, New York.

Balakrishnan, N., Lai, C.D. (2009). Continuous Bivariate Distributions. Springer-Verlag, New York.

Bourguignon, M., Cordeiro, G.M. (2016). New results on the Ristic- Balakrishnan family of distributions.
Communications in Statistics- Theory and Methods. DOI: 10.1080/03610926.2014.972573.

Gl\"{a}nzel, W. (1987). A characterization theorem based on truncated moments and its application to
some distribution families. In Mathematical Statistics and Probability Theory (Bad Tatzmannsdorf, 1986),
Vol. B (p. 75-84). Dordrecht: Reidel.

Hamedani, G.G. (2010). Characterizations of univariate continuous distributions based on truncated
moments of functions of order statistics. Studia Scientiarum Mathematicarum Hungarica, 47, 462-468.

Kotz, S., Balakrishnan, N., and Johnson, N.L. (2000). Continuous Multivariate Distributions, Volume
1, Models and Applications, 2nd Edition. John Wiley, New York.
Ramos, H. M., Ollero, J. and Sordo, M. A. (2000). A sufficient condition for Generalized Lorenz order.
Journal of Economic Theory, 90, 286-292.

Ristic, M. and Balakrishnan, N. (2012). The gamma-exponentiated exponential distribution. Journal of
Statistical Computation and Simulation, 82, 1191-1206.

Ristic, M. and Balakrishnan, N. (2016). Multivariate families of gamma-generated distributions with
finite or infinite support above or below the diagonal. Journal of Multivariate Analysis, 143, 194-207.

\end{document}